\documentclass[aps,prl,reprint,showpacs,floatfix,amsmath,amssymb,superscriptaddress]{revtex4-1}
\usepackage{graphicx}

\usepackage{amsmath}
\usepackage{physics}
\usepackage{bm}
\usepackage{here}
\begin{document}

\newcommand{\Yb}[1]{$^{#1}\mathrm{Yb}$}
\newcommand{\state}[3]{$^{#1}\mathrm{#2}_{#3}$}

\title{Observation of spin-exchange dynamics between itinerant and localized \Yb{171} atoms}
\author{Koki Ono}
\altaffiliation{Electronic address: koukiono3@yagura.scphys.kyoto-u.ac.jp}
\affiliation{Department of Physics, Graduate School of Science, Kyoto University, Kyoto 606-8502, Japan}
\author{Yoshiki Amano}
\affiliation{Department of Physics, Graduate School of Science, Kyoto University, Kyoto 606-8502, Japan}
\author{Toshiya Higomoto}
\affiliation{Department of Physics, Graduate School of Science, Kyoto University, Kyoto 606-8502, Japan}
\author{Yugo Saito}
\affiliation{Department of Physics, Graduate School of Science, Kyoto University, Kyoto 606-8502, Japan}
\author{Yoshiro Takahashi}
\affiliation{Department of Physics, Graduate School of Science, Kyoto University, Kyoto 606-8502, Japan}
\date{\today}
\begin{abstract}
We report on the observation of the spin-exchange dynamics of \Yb{171} atoms in the ground state \state{1}{S}{0} and in the metastable state \state{3}{P}{0}. We implement the mixed-dimensional two-orbital system using a near-resonant and magic-wavelength optical lattices, where the \state{1}{S}{0} and \state{3}{P}{0} atoms are itinerant in a one-dimensional tube and localized in three dimensions, respectively. By exploiting an optical Stern-Gerlach method, we observe the spin depolarization of the \state{1}{S}{0} atoms induced by the spin-exchange interaction with the \state{3}{P}{0} atom. Our work could open the way to the quantum simulation of the Kondo effect.

\end{abstract}
\maketitle

\section{I. Introduction}
Strongly correlated systems with orbital degrees of freedom exhibit interesting phenomena, with the Kondo effect \cite{Kondo1964}, which is the many-body phenomenon arising from an antiferromagnetic interaction between a conduction electron and a localized magnetic moment, as a prominent example. It was originally studied in the context of the enhancement of the resistivity in magnetic alloys at low temperature, and it is now a ubiquitous problem in condensed matter physics. Also, the Kondo lattice model, where the localized spins are aligned periodically, is a paradigmatic model of a heavy fermion system. Its phase diagram, called the Doniach phase diagram \cite{Doniach1977}, contains the paramagnetic phase due to the Kondo screening in the strong coupling regime and the Ruderman-Kittel-Kasuya-Yoshida (RKKY) ordered phase in the weak coupling regime. 

Although the Kondo effect has been intensively studied in solid-state and mesoscopic systems, by exploiting its novel possibilities in the control of system parameters and the detection, ultracold atomic gases in an optical lattice allow one to study the Kondo system in a unique manner, which is challenging to investigate in other systems \cite{Bloch2012}. In particular, alkaline-earth-like atoms have received much attention due to the presence of the metastable states \state{3}{P}{0} and \state{3}{P}{2} as well as the ground state \state{1}{S}{0}. Taking advantage of the long-lived states, the quantum simulator with orbital degrees of freedom using the \state{1}{S}{0} and \state{3}{P}{0} or \state{3}{P}{2} atoms has been proposed, and the Kondo system using the two-orbital system has been studied theoretically \cite{Gorshkov2010,Foss2010,Nakagawa2015,Zhang2016,Kanasz2018,Nakagawa2018,Kuzmenko2018,Goto2019}. In order to implement the Kondo system with cold atoms, an antiferromagnetic spin-exchange interaction between mobile and immobile atoms is required. The clock transition spectroscopy in the state-independent optical lattice, called the magic-wavelength optical lattice, reveals that, in contrast to a ferromagnetic coupling of \Yb{173} and {$^{87}\mathrm{Sr}$ \cite{Cappellini2014,Scazza2014,Zhang2014}, the fermionic isotope of \Yb{171} has an antiferromagnetic coupling between the \state{1}{S}{0} atom and the \state{3}{P}{0} atom \cite{Ono2019}. This suggests that the two-orbital system using \Yb{171} is a promising natural candidate for the quantum simulator of the Kondo effect in contrast to the tuning of the spin-exchange coupling via confinement-induced resonances \cite{Riegger2018}. Another feature of \Yb{171} is the weak interatomic interaction of \state{1}{S}{0} atoms, suggesting that the \state{1}{S}{0} atoms in an optical lattice can be described as a non-interacting metallic state, which is suitable for the exploration of the Kondo physics. Motivated by these unique properties of the two-orbital system using \Yb{171}, the numerical simulation of the dipole oscillation of the \state{1}{S}{0} atoms in the presence of the localized \state{3}{P}{0} atom is performed \cite{Goto2019}, showing that the Kondo effect manifests itself in such a way that the center-of-mass motion of \state{1}{S}{0} atoms is suppressed as the temperature is lowered due to the antiferromagnetic spin-exchange interaction.

In this letter, we report on the observation of the spin-exchange dynamics between \Yb{171} atoms in the ground state $\ket{g}=\ket{^1\mathrm{S}_0}$ and in the metastable state $\ket{e} = \ket{^3\mathrm{P}_0}$. Using a two-orbital lattice system consisting of a two-dimensional (2D) magic-wavelength optical lattice and a 1D near-resonant optical lattice giving strong confinement to the $\ket{e}$ atom alone and no net effect to the $\ket{g}$ atom, the quasi (0+1)D system is implemented, where the $\ket{g}$ atom behaves as the quasi 1D free fermion interacting with the $\ket{e}$ atom mimicking a localized magnetic moment. By exploiting the optical Stern-Gerlach method, we observe the relaxation of the spin polarization caused by the interorbital spin-exchange process and the suppression of the spin depolarization in a high magnetic field. The rate of spin-exchange dynamics is also controlled by the excited-state population. These observations are an important first step towards the quantum simulation of the Kondo effect. 

\begin{figure*}
\includegraphics[width=1\linewidth]{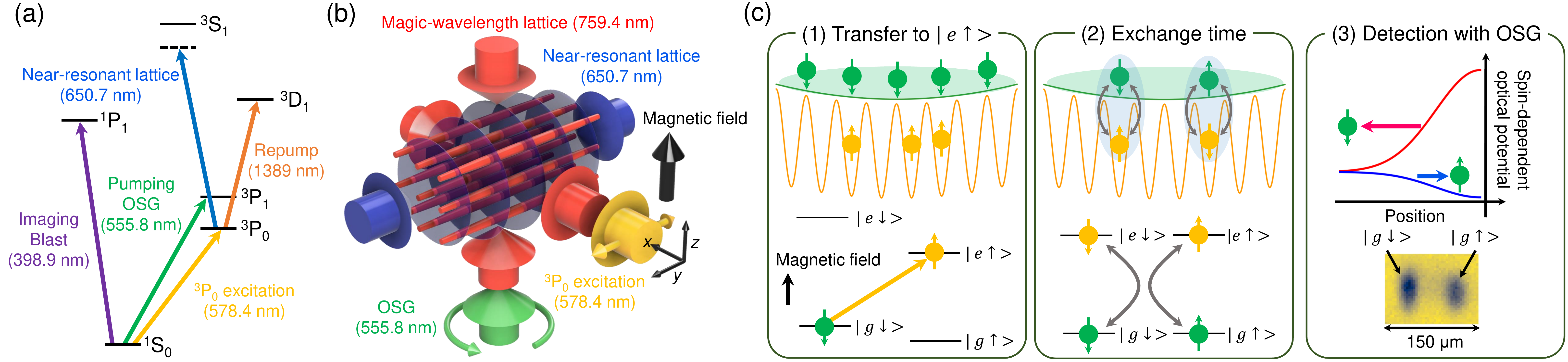}
\caption{Schematic diagram of experiment. (a) Relevant energy diagram of Yb atom. (b) Schematic illustration of beam configuration. The polarization of the clock excitation light is perpendicular to the quantization axis defined by the magnetic
field, and it amounts to the equal mixture of $\sigma_+$ and $\sigma_-$ polarization. The circularly polarized OSG light propagates along the quantization axis. (c) Schematic illustration of experimental procedure. (1) Initially, some fraction of the atoms in the $\ket{g\downarrow}$ state (a green ball) are excited to the $\ket{e\uparrow}$ state (a yellow ball) in a magnetic field of 30 G. The upper figure shows the schematic representation of the optical lattice potentials for the $\ket{g}$ atom (a green curve) and the $\ket{e}$ atom (a yellow curve). (2) After the excitation, a magnetic field is lowered to 0.5 G, and the spin-exchange dynamics is started. (3) After the hold time, the population of the atoms in the $\ket{g\uparrow}$ and in the $\ket{g\downarrow}$ is detected with an OSG technique. The upper figure shows the spin-dependent optical gradient potential to spatially separate the $\ket{g\uparrow}$ atom and the $\ket{g\downarrow}$ atom. The lower figure shows a typical example of the simultaneous observation of both spin states in the false color ToF image of the \Yb{171} gas in the $\ket{g}$ state subjected to the OSG light.}
\label{fig1}
\end{figure*}

\section{II. Methods}
We first explain how we implement the quasi-(0+1)D system using a near-resonant optical lattice. An optical dipole potential $V(\vb*{r})$ is proportional to the laser intensity $I(\vb*{r})$:
\begin{equation}
V(\vb*{r})=-\frac{1}{4}\alpha I(\vb*{r}).
\label{eq1}
\end{equation}
The coefficient $\alpha$ is called polarizability:
\begin{equation}
\alpha = \sum_i \frac{6\pi c^2}{\omega_{i}^3}\left(\frac{\Gamma_i}{\omega_{i}-\omega}+\frac{\Gamma_i}{\omega_{i}+\omega}\right),
\label{eq2}
\end{equation}
where $\omega$ is the laser angular frequency and $c$ is the speed of light. Here $\omega_i$ and $\Gamma_i$ correspond to the resonant angular frequency and the natural linewidth of the $i$th-state, respectively. The wavelength of the near-resonant optical lattice is chosen to be 650.7 nm, which is close to the \state{3}{P}{0}--\state{3}{S}{1} transition wavelength of 649.1 nm, resulting in the large polarizability for the $\ket{e}$ atom $\alpha_e$ (see Fig.~1(a) for relevant energy levels). Using the Eq.~(\ref{eq2}), the polarizability is obtained as $\alpha_e/h=$1.4 kHz/(mW/cm$^2$), $h$ being the Planck constant. In this calculation, we assume that the \state{3}{P}{0}--\state{3}{S}{1} transition makes the dominant contribution and the other transitions are negligible. Similarly, the polarizability for the $\ket{g}$ atom $\alpha_g$ due to the 650.7 nm light is also calculated as $\alpha_g/h=39$ Hz/(mW/cm$^2$). As a result, the near-resonant optical lattice has the large polarizability ratio $\alpha_e/\alpha_g=36$, while a similar experiment in Ref. \cite{Riegger2018} uses a state-dependent lattice with  $\alpha_e/\alpha_g=3.3$. In our experiment, the $\ket{e}$ atom is deeply confined by the near-resonant lattice while the lattice potential is regarded as the continuum system for the $\ket{g}$ atom. A natural concern of using near-resonant light is the possibly non-negligible photon scattering loss rate $\gamma_\mathrm{sc}$ of the $\ket{e}$ atoms. By using narrow-linewidth band-pass filters with less than 0.1~nm to suppress the 649.1 nm resonant frequency component in the single-mode narrow-linewidth 650.7 nm laser, we obtain the loss rate of the $\ket{e}$ atom in the absence of the $\ket{g}$ atom $\gamma_\mathrm{sc}$ = 5.0 Hz. Although this is larger than the theoretically estimated value of 0.59 Hz obtained by assuming only the off-resonant excitation, the corresponding lifetime is long enough to clearly observe the spin-exchange dynamics (see III. RESULTS). A state-independent optical lattice is obtained with the magic wavelength of 759.4 nm.

Figure \ref{fig1}(b) illustrates the schematic diagram of the beam configuration. The 2D array of the tube traps is produced using the 2D magic-wavelength lattice ($x$ and $z$), and the 1D near-resonant optical lattice is superimposed along the axis of the tubes ($y$). As a result, the $\ket{e}$ atom is localized by the 3D confinement while the $\ket{g}$ atom is mobile along the $y$ direction in the tube potential. The maximum potential depth for the $\ket{e}$ atom due to the near-resonant lattice amounts to 27$E_\mathrm{R}$, with $E_\mathrm{R}=k_\mathrm{B}\times96$~nK being the recoil energy for the magic wavelength. Here $k_\mathrm{B}$ is the Boltzmann constant. The corresponding trap frequency of the lattice site at the trap center is 24 kHz, and the residual harmonic trap frequency due to the Gaussian beam shape is estimated as 24 Hz from the beam radius.

The near-resonant lattice depth is calibrated using diffraction of the $\ket{e}$ atoms by a pulsed optical lattice technique with the near-resonant lattice \cite{Denschlag2002}. We use the \Yb{171} atom for the calibration of the lattice depth since the bosonic isotopes would suffer from the severe inelastic loss in the \state{3}{P}{0} states, in addition to another technical merit that the Rabi frequency of the clock transition for the fermionic isotopes is larger than that for bosonic isotopes. After the excitation to the $\ket{e}$ state in the 3D magic-wavelength lattice, the remaining atoms in the $\ket{g}$ state are blasted with the resonant light with the \state{1}{S}{0}--\state{1}{P}{1} transition. Then the magic-wavelength lattice potential along the $y$ axis is ramped down in 1 ms, and the pulsed lattice is irradiated along the $y$ axis immediately after switching off the remaining magic-wavelength lattice potentials along the $x$ and $z$ axes. During the time-of-flight (ToF), the atoms are repumped into the $\ket{g}$ state using the resonant light with the \state{3}{P}{0}--\state{3}{D}{1} transition, and the diffraction pattern is probed by absorption imaging with the \state{1}{S}{0}--\state{1}{P}{1} transition. From the oscillatory behavior of the diffraction pattern, we can calibrate the near-resonant lattice depth. 

Our experiments start with the preparation of the quantum degenerate gas of \Yb{171} using the sympathetic evaporative cooling with \Yb{173} \cite{Ono2019}. During the evaporative cooling, the optical pumping into the $\ket{g\downarrow}$ state is performed with the \state{1}{S}{0}--\state{3}{P}{1}($F'=1/2$) transition, where $\ket{\uparrow}=\ket{m_F=+1/2}$, $\ket{\downarrow}=\ket{m_F=-1/2}$ denote the projections of the nuclear spin $F$ onto the quantization axis defined by a magnetic field. The number of atoms $N$ and the temperature scaled by the Fermi temperature $T/T_{\mathrm{F}}$ are $N \simeq 2\times10^4$ and $T/T_{\mathrm{F}}\simeq0.3$, respectively. After the removal of \Yb{173} atoms using the resonant light associated with the \state{1}{S}{0}--\state{3}{P}{1}$(F'=7/2$) transition, the atoms are loaded into the optical lattices, where the initial depths of the magic-wavelength optical lattice and the near-resonant optical lattice for the $\ket{e}$ atom are set to $30E_{\mathrm{R}}$ and $6.8E_{\mathrm{R}}$, respectively. Figure \ref{fig1}(c) shows the experimental procedure after loading atoms into an optical lattice. (1) Some fraction of the atoms are coherently transferred to the $\ket{e\uparrow}$ state in a magnetic field of 30 G by a stabilized clock laser \cite{Takata2019} with a typical linewidth of a few Hz. To localize the $\ket{e}$ atom, the near-resonant optical lattice is then ramped up to 27$E_\mathrm{R}$ in 1 ms, which is longer than the inverse of the lattice-site trap frequency, regarded as an adiabatic ramp. To reduce the spatial inhomogeneity of the clock transition frequency due to the residual harmonic trap created by the near-resonant optical lattice, the $\ket{g}$ atoms are coherently transferred to the $\ket{e}$ state in the shallower optical lattice. A moderate lattice depth is required for the sideband-resolved excitation, on the other hand, which allows one to prepare $\ket{e}$ atoms in the lowest band. We thus perform the clock excitation with the lattice depth of 6.8$E_\mathrm{R}$. The hopping energies between the nearest neighbor tubes $J_\perp$ and between the adjacent sites of the near-resonant optical lattice $J_y$ are estimated as $J_\perp=h\times1.0$ Hz and  $J_y=h\times7.0$ Hz for the $\ket{e}$ atom, indicating that the hopping energy is negligible within the experimentally relevant timescales. (2) After the clock excitation, the magnetic field is rapidly lowered to 0.5 G with about 3 ms settling time, and the spin-exchange dynamics is initiated. (3) After the hold time, the spin polarization of the $\ket{g}$ atoms is detected with the optical Stern-Gerlach technique (OSG) \cite{Taie2010}, which enables one to separately observe the atoms in the $\ket{g\uparrow}$ and $\ket{g\downarrow}$ states using a spin-dependent optical potential gradient. The OSG light is blue-detuned by 875 MHz from the \state{1}{S}{0}--\state{3}{P}{1}($F'=1/2$) transition.

\section{III. Results}

\begin{figure}
\includegraphics[width=0.95\linewidth]{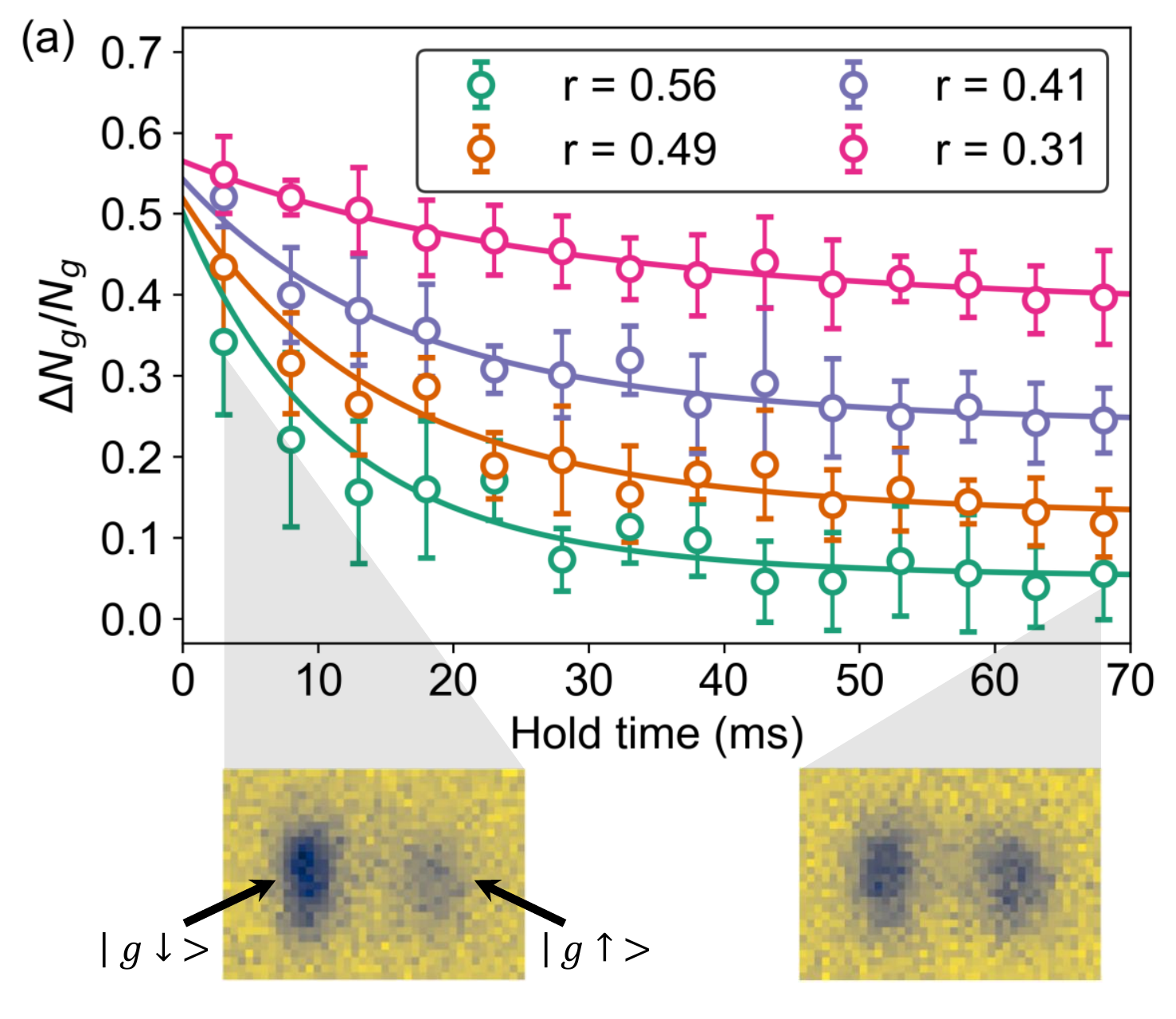}
\includegraphics[width=0.8\linewidth]{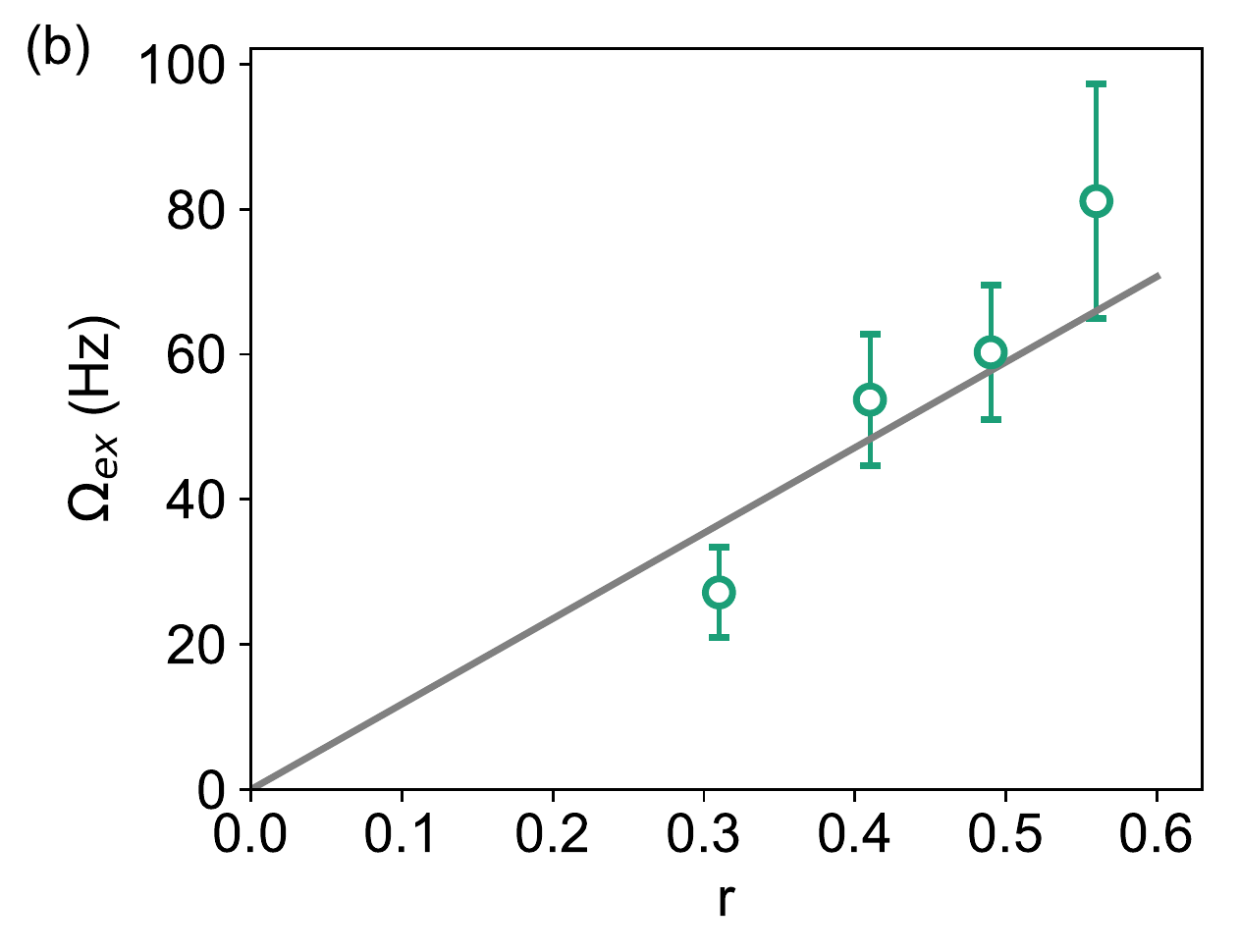}
\caption{Observation of spin-exchange dynamics. (a) Time evolution of the spin polarization of the $\ket{g}$ atoms $\Delta N_g/N_g$ with the different excitation rates to the $\ket{e}$ state: $r=0.56$, $0.49$, $0.41$, and $0.31$. Error bars show the standard deviations of the mean values obtained by averaging ten measurements. Solid lines represent fits to the data with the Eqs~(\ref{eq3})-(\ref{eq6}).  False color time-of-flight images of the $\ket{g}$ atoms after the spin-exchange dynamics with $r=0.56$ are shown. The left and right figures correspond to the hold time of 3 ms and 68 ms, respectively. (b) Spin-exchange rate $\Omega_\mathrm{ex}$ as a function of the excitation rate. Error bars are 1$\sigma$ confidence intervals of the data fits. The solid line represents linear fits to the data.}
\label{fig2}
\end{figure}

Figure \ref{fig2}(a) shows the time evolution of the spin polarization of the $\ket{g}$ atoms, defined as $\Delta N_g/N_{g}$. Here $\Delta N_g$ denotes the atom number difference between the $\ket{g\uparrow}$ and $\ket{g\downarrow}$ states, and $N_g$ is the total number of the $\ket{g}$ atoms. The result clearly shows the spin depolarization due to the spin-exchange interaction with $\ket{e}$ atoms. We note that we did not observe the depolarization in the case of no $\ket{e}$ atoms. Also, the relaxation rate of the spin polarization is controlled by the clock excitation rate $r$, which is associated with the number of $\ket{e}$ atoms. The spin polarization less than unity at the initial time could be ascribed to the imperfect optical pumping and the photon-scattering of the OSG light. It is noted that the remaining unwanted spin component is not removed after the optical pumping.

The observed relaxation dynamics is quantitatively analyzed with the following two-body rate equations \cite{Scazza2014}:

\begin{align}
\dot{p}_{g\uparrow}(t) =& \Omega_\mathrm{ex}(p_{g\downarrow}(t)p_{e\uparrow}(t)-p_{e\downarrow}(t)p_{g\uparrow}(t)) \nonumber \\  
&-\Gamma_{eg}p_{g\uparrow}(t)(p_{e\uparrow}(t)+p_{e\downarrow}(t)) \nonumber \\ 
&+ \frac{\gamma_\mathrm{sc}}{2}(p_{e\uparrow}(t)+p_{e\downarrow}(t)),
\label{eq3}
\end{align}
\begin{align}
\dot{p}_{g\downarrow}(t) =& \Omega_\mathrm{ex}(p_{e\downarrow}(t)p_{g\uparrow}(t)-p_{g\downarrow}(t)p_{e\uparrow}(t)) \nonumber \\  
&-\Gamma_{eg}p_{g\downarrow}(t)(p_{e\uparrow}(t)+p_{e\downarrow}(t)) \nonumber \\ 
&+ \frac{\gamma_\mathrm{sc}}{2}(p_{e\uparrow}(t)+p_{e\downarrow}(t)),
\label{eq4}
\end{align}
\begin{align}
\dot{p}_{e\uparrow}(t) =& \Omega_\mathrm{ex}(p_{e\downarrow}(t)p_{g\uparrow}(t)-p_{g\downarrow}(t)p_{e\uparrow}(t)) \nonumber \\  
&-\Gamma_{eg}p_{e\uparrow}(t)(p_{g\uparrow}(t)+p_{g\downarrow}(t)) \nonumber \\ 
&- \gamma_\mathrm{sc}p_{e\uparrow}(t),
\label{eq5}
\end{align}
\begin{align}
\dot{p}_{e\downarrow}(t) =& \Omega_\mathrm{ex}(p_{g\downarrow}(t)p_{e\uparrow}(t)-p_{e\downarrow}(t)p_{g\uparrow}(t)) \nonumber \\  
&-\Gamma_{eg}p_{e\downarrow}(t)(p_{g\uparrow}(t)+p_{g\downarrow}(t)) \nonumber \\ 
&- \gamma_\mathrm{sc}p_{e\downarrow}(t).
\label{eq6}
\end{align}
Here $p_{\alpha\sigma}(t)=\bar{n}_{\alpha\sigma}(t)/\bar{n}_0$ denotes the relative population of the atom in the $\ket{\alpha\sigma}$ state ($\alpha=g,e$, $\sigma = \uparrow, \downarrow$), where $\bar{n}_{\alpha\sigma}$ and $\bar{n}_0$ denote the mean density of the atom in the $\ket{\alpha\sigma}$ state and the mean density of the total atoms in the initial state, respectively. Also, $\Omega_\mathrm{ex}$ and $\Gamma_{eg}$ correspond to the spin-exchange rate and the two-body loss rate between the $\ket{g}$ atom and the $\ket{e}$ atom, respectively, and they are propotional to $\bar{n}_0$. In addition, $\gamma_\mathrm{sc}$ is the one-body loss rate of the $\ket{e}$ atom.
We assume that the inelastic collision between the $\ket{e}$ atoms is ignored since the hopping rates $J_\perp$ and $J_y$ are much smaller than the spin depolarization rate. 
In addition, $\Gamma_{eg}$ is assumed to be independent of the spin state and is estimated from the measurement of the lifetime of the $\ket{e}$ atom during the spin-exchange dynamics, resulting in $\Gamma_{eg}=10$ Hz. 
On the other hand, using the inelastic loss-rate coefficient $\beta_{eg\pm} \leq 2.6(3)\times 10^{-16} $ cm$^3$/s obtained by the measurement of the lifetimes of the $\ket{eg^+}$ and $\ket{eg^-}$ states in the 3D magic-wavelength optical lattice \cite{Bettermann2020}, the two-body loss rate is calculated as $2.3\times10^{-2}$ Hz, where $\ket{eg^+}$ and $\ket{eg^-}$ correspond to the spin-singlet state and the spin-triplet state, respectively. 
Although the origin of the discrepancy is not known, here we note that the analysis using Eqs.~(\ref{eq3})-(\ref{eq6}) does not depend sensitively on the value of $\Gamma_{eg}$, due to the existence of photon scattering loss $\gamma_\mathrm{sc} = 5$ Hz, and in fact $\Gamma_{eg} =10$ Hz and $\Gamma_{eg} = 0$ give almost the same results. In the following analysis, we adopt $\Gamma_\mathrm{eg} =10$ Hz.
It should be noted that the two-body loss between the $\ket{e}$ atoms via a tunneling process is suppressed owing to the on-site repulsive interaction and the two-body dissipation via a quantum Zeno effect \cite{Syassen2008,Tomita2019}. In our experiment, this effective loss rate is estimated as 0.28 Hz, suggesting that the inelastic loss between the $\ket{e}$ atoms would not occur during the spin-exchange dynamics. 
Solid lines in Fig.~\ref{fig2}(a) represent the fits to the data using the two-body rate equations (\ref{eq3})-(\ref{eq6}) by treating $\Omega_\mathrm{ex}$ as a free parameter. Figure \ref{fig2}(b) shows the spin-exchange rate obtained from the data fits in Fig.~\ref{fig2}(a) as a function of the  excitation rate to the $\ket{e}$ state, exhibiting an enhancement of the spin-exchange rate with the increase of the number of atoms in the $\ket{e}$ state. 
The linear dependence of the spin-exchange rate $\Omega_\mathrm{ex}$ on the excitation rate $r$ is expected when no correlation between the $\ket{e}$ atoms is considered. The validity of this assumption is related to the characteristic energy of the RKKY interaction $V_\mathrm{ex}^2/\epsilon_\mathrm{F}$ \cite{Tsunetsugu1997}, where $V_\mathrm{ex}$ and $\epsilon_\mathrm{F}$ are the spin-exchange energy and the Fermi energy, respectively. In our experiment, this is estimated as $k_\mathrm{B}\times1.4$~nK, which is much smaller than the atomic temperature. Thus, the RKKY correlation is negligible. In addition, this linearity is expected for a small excitation rate, and the investigation of the spin-exchange dynamics with a higher excitation rate will be an interesting future study.

\begin{figure}
\includegraphics[width=0.85\linewidth]{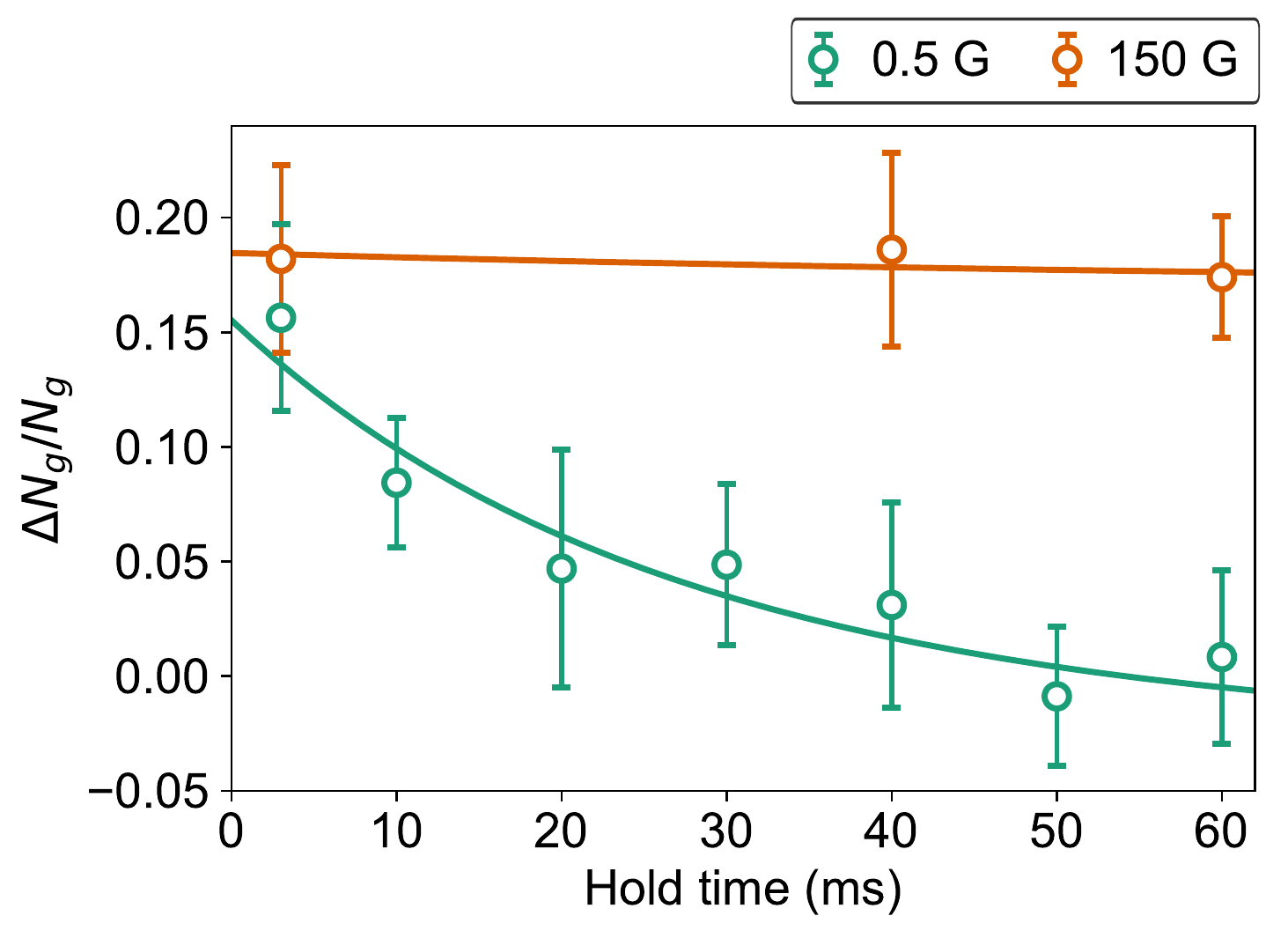}
\caption{Magnetic-field dependence of spin-exchange dynamics with the \state{3}{P}{0} excitation rate $r=0.56$. Error bars show the standard deviations of the mean values obtained by averaging ten measurements. In this experiment, the initial state of the exchange dynamics is prepared by the excitation $\ket{g\uparrow}\to\ket{e\uparrow}$ of the spin-balanced \Yb{171} atoms using $\pi$-polarized light since this scheme results in the reduced uncertainty.}
\label{fig3}
\end{figure}

Also, we investigate the magnetic-field dependence of the spin-exchange dynamics. The interorbital spin-exchange interaction energy $V_\mathrm{ex}$ can be estimated as 
\begin{equation}
V_\mathrm{ex} = \frac{4\pi\hbar^2}{m}\frac{a_{eg}^+-a_{eg}^-}{2}\int d^3\vb*{r}n_g(\vb*{r})\abs{\psi_e(\vb*{r})}^2,
\label{eq8}
\end{equation}
where $m$ denotes the mass of the atom and $\psi_e(\vb*{r})$ is the single-particle wavefunction of the $\ket{e}$ atom. The spin-exchange is characterized by the difference between the spin-singlet scattering length $a_{eg}^+=240(4)a_0$ and the spin-triplet scattering length $a_{eg}^- = 389(4)a_0$ \cite{Bettermann2020}, resulting in an antiferromagnetic coupling $V_\mathrm{ex}<0$. Here $a_0$ denotes the Bohr radius. In the central tube, $V_\mathrm{ex}$ is estimated to be $V_\mathrm{ex}/h = -0.25$ kHz.
On the other hand, the differential Zeeman shift between the $\ket{g\uparrow}$ state and the $\ket{e\uparrow}$ state amounts to $-200.0(6)$ Hz/G \cite{Ono2019}. Figure \ref{fig3} shows a comparison between the spin relaxation dynamics in a magnetic field of 0.5 G and that in a magnetic field of 150 G, where the Zeeman energy is two orders of magnitude larger than $V_\mathrm{ex}$. The result shows that the spin depolarization dynamics is frozen in a higher magnetic field, which is consistent with the fact that the spin-exchange process is energetically costly.

\section{IV. Conclusions}
In conclusion, we successfully realized the quasi-1D fermion system in the presence of the immobile spin using the 2D magic-wavelength optical lattice and the 1D near-resonant optical lattice. Using this system, the spin-exchange dynamics between the itinerant $\ket{g}$ atom and the localized $\ket{e}$ atom is observed. Our works can pave the way to the quantum simulation of the Kondo effect. Although the near-resonant lattice causes the one-body loss of the $\ket{e}$ atom, which is detrimental to the exploration of many-body physics, the scattering rate $\gamma_\mathrm{sc}$ will be reduced by using a far-detuned laser source with, for example, a wavelength of 652 nm. It will be an interesting in future work to compare the observed spin-exchange rates with theoretical calculations \cite{Cheng2017,Zhang2018,GotoPrivCom}. The Kondo effect manifests itself as a many-body singlet state, resulting in the screening of a localized spin by itinerant fermions, which is called Kondo screening. The screening cloud of itinerant fermions could be observed in the two-orbital system using a Yb quantum gas microscope \cite{Yamamoto2016,Miranda2015}, as in a quantum dot system \cite{Borzenets2020}. In addition, it is interesting to detect the spin correlation between the excited-state atoms as a signature of the RKKY interaction \cite{Gorshkov2010}.  

\section{acknowledgments}
\begin{acknowledgments}
We thank Ippei Danshita, Shimpei Goto for fruitful discussions. KO acknowledges support from the JSPS (KAKENHI grant number 19J11413). The experimental work was supported by the Grant-in-Aid for Scientific Research of JSPS(Nos.~JP17H06138, JP18H05405, and JP18H05228), the Impulsing Paradigm Change through Disruptive Technologies (ImPACT) program, JST CREST (No.~JP-MJCR1673), and MEXT Quantum Leap Flagship Program (MEXT Q-LEAP) Grant No.~JPMXS0118069021.
\end{acknowledgments}


\begin{thebibliography}{29}%
\makeatletter
\providecommand \@ifxundefined [1]{%
 \@ifx{#1\undefined}
}%
\providecommand \@ifnum [1]{%
 \ifnum #1\expandafter \@firstoftwo
 \else \expandafter \@secondoftwo
 \fi
}%
\providecommand \@ifx [1]{%
 \ifx #1\expandafter \@firstoftwo
 \else \expandafter \@secondoftwo
 \fi
}%
\providecommand \natexlab [1]{#1}%
\providecommand \enquote  [1]{``#1''}%
\providecommand \bibnamefont  [1]{#1}%
\providecommand \bibfnamefont [1]{#1}%
\providecommand \citenamefont [1]{#1}%
\providecommand \href@noop [0]{\@secondoftwo}%
\providecommand \href [0]{\begingroup \@sanitize@url \@href}%
\providecommand \@href[1]{\@@startlink{#1}\@@href}%
\providecommand \@@href[1]{\endgroup#1\@@endlink}%
\providecommand \@sanitize@url [0]{\catcode `\\12\catcode `\$12\catcode
  `\&12\catcode `\#12\catcode `\^12\catcode `\_12\catcode `\%12\relax}%
\providecommand \@@startlink[1]{}%
\providecommand \@@endlink[0]{}%
\providecommand \url  [0]{\begingroup\@sanitize@url \@url }%
\providecommand \@url [1]{\endgroup\@href {#1}{\urlprefix }}%
\providecommand \urlprefix  [0]{URL }%
\providecommand \Eprint [0]{\href }%
\providecommand \doibase [0]{http://dx.doi.org/}%
\providecommand \selectlanguage [0]{\@gobble}%
\providecommand \bibinfo  [0]{\@secondoftwo}%
\providecommand \bibfield  [0]{\@secondoftwo}%
\providecommand \translation [1]{[#1]}%
\providecommand \BibitemOpen [0]{}%
\providecommand \bibitemStop [0]{}%
\providecommand \bibitemNoStop [0]{.\EOS\space}%
\providecommand \EOS [0]{\spacefactor3000\relax}%
\providecommand \BibitemShut  [1]{\csname bibitem#1\endcsname}%
\let\auto@bib@innerbib\@empty
\bibitem [{\citenamefont {Kondo}(1964)}]{Kondo1964}%
  \BibitemOpen
  \bibfield  {author} {\bibinfo {author} {\bibfnamefont {J.}~\bibnamefont
  {Kondo}},\ }\href {\doibase 10.1143/PTP.32.37} {\bibfield  {journal}
  {\bibinfo  {journal} {Progress of Theoretical Physics}\ }\textbf {\bibinfo
  {volume} {32}},\ \bibinfo {pages} {37} (\bibinfo {year} {1964})}\BibitemShut
  {NoStop}%
\bibitem [{\citenamefont {Doniach}(1977)}]{Doniach1977}%
  \BibitemOpen
  \bibfield  {author} {\bibinfo {author} {\bibfnamefont {S.}~\bibnamefont
  {Doniach}},\ }\href {\doibase https://doi.org/10.1016/0378-4363(77)90190-5}
  {\bibfield  {journal} {\bibinfo  {journal} {Physica B+ C}\ }\textbf {\bibinfo
  {volume} {91}},\ \bibinfo {pages} {231} (\bibinfo {year} {1977})}\BibitemShut
  {NoStop}%
\bibitem [{\citenamefont {Bloch}\ \emph {et~al.}(2012)\citenamefont {Bloch},
  \citenamefont {Dalibard},\ and\ \citenamefont {Nascimbene}}]{Bloch2012}%
  \BibitemOpen
  \bibfield  {author} {\bibinfo {author} {\bibfnamefont {I.}~\bibnamefont
  {Bloch}}, \bibinfo {author} {\bibfnamefont {J.}~\bibnamefont {Dalibard}}, \
  and\ \bibinfo {author} {\bibfnamefont {S.}~\bibnamefont {Nascimbene}},\
  }\href {\doibase 10.1038/nphys2259} {\bibfield  {journal} {\bibinfo
  {journal} {Nature Physics}\ }\textbf {\bibinfo {volume} {8}},\ \bibinfo
  {pages} {267} (\bibinfo {year} {2012})}\BibitemShut {NoStop}%
\bibitem [{\citenamefont {Gorshkov}\ \emph {et~al.}(2010)\citenamefont
  {Gorshkov}, \citenamefont {Hermele}, \citenamefont {Gurarie}, \citenamefont
  {Xu}, \citenamefont {Julienne}, \citenamefont {Ye}, \citenamefont {Zoller},
  \citenamefont {Demler}, \citenamefont {Lukin},\ and\ \citenamefont
  {Rey}}]{Gorshkov2010}%
  \BibitemOpen
  \bibfield  {author} {\bibinfo {author} {\bibfnamefont {A.~V.}\ \bibnamefont
  {Gorshkov}}, \bibinfo {author} {\bibfnamefont {M.}~\bibnamefont {Hermele}},
  \bibinfo {author} {\bibfnamefont {V.}~\bibnamefont {Gurarie}}, \bibinfo
  {author} {\bibfnamefont {C.}~\bibnamefont {Xu}}, \bibinfo {author}
  {\bibfnamefont {P.~S.}\ \bibnamefont {Julienne}}, \bibinfo {author}
  {\bibfnamefont {J.}~\bibnamefont {Ye}}, \bibinfo {author} {\bibfnamefont
  {P.}~\bibnamefont {Zoller}}, \bibinfo {author} {\bibfnamefont
  {E.}~\bibnamefont {Demler}}, \bibinfo {author} {\bibfnamefont {M.~D.}\
  \bibnamefont {Lukin}}, \ and\ \bibinfo {author} {\bibfnamefont
  {A.}~\bibnamefont {Rey}},\ }\href {\doibase 10.1038/nphys1535} {\bibfield
  {journal} {\bibinfo  {journal} {Nature physics}\ }\textbf {\bibinfo {volume}
  {6}},\ \bibinfo {pages} {289} (\bibinfo {year} {2010})}\BibitemShut {NoStop}%
\bibitem [{\citenamefont {Foss-Feig}\ \emph {et~al.}(2010)\citenamefont
  {Foss-Feig}, \citenamefont {Hermele},\ and\ \citenamefont {Rey}}]{Foss2010}%
  \BibitemOpen
  \bibfield  {author} {\bibinfo {author} {\bibfnamefont {M.}~\bibnamefont
  {Foss-Feig}}, \bibinfo {author} {\bibfnamefont {M.}~\bibnamefont {Hermele}},
  \ and\ \bibinfo {author} {\bibfnamefont {A.~M.}\ \bibnamefont {Rey}},\ }\href
  {\doibase 10.1103/PhysRevA.81.051603} {\bibfield  {journal} {\bibinfo
  {journal} {Phys. Rev. A}\ }\textbf {\bibinfo {volume} {81}},\ \bibinfo
  {pages} {051603(R)} (\bibinfo {year} {2010})}\BibitemShut {NoStop}%
\bibitem [{\citenamefont {Nakagawa}\ and\ \citenamefont
  {Kawakami}(2015)}]{Nakagawa2015}%
  \BibitemOpen
  \bibfield  {author} {\bibinfo {author} {\bibfnamefont {M.}~\bibnamefont
  {Nakagawa}}\ and\ \bibinfo {author} {\bibfnamefont {N.}~\bibnamefont
  {Kawakami}},\ }\href {\doibase 10.1103/PhysRevLett.115.165303} {\bibfield
  {journal} {\bibinfo  {journal} {Physical review letters}\ }\textbf {\bibinfo
  {volume} {115}},\ \bibinfo {pages} {165303} (\bibinfo {year}
  {2015})}\BibitemShut {NoStop}%
\bibitem [{\citenamefont {Zhang}\ \emph {et~al.}(2016)\citenamefont {Zhang},
  \citenamefont {Zhang}, \citenamefont {Cheng}, \citenamefont {Chen},
  \citenamefont {Zhang},\ and\ \citenamefont {Zhai}}]{Zhang2016}%
  \BibitemOpen
  \bibfield  {author} {\bibinfo {author} {\bibfnamefont {R.}~\bibnamefont
  {Zhang}}, \bibinfo {author} {\bibfnamefont {D.}~\bibnamefont {Zhang}},
  \bibinfo {author} {\bibfnamefont {Y.}~\bibnamefont {Cheng}}, \bibinfo
  {author} {\bibfnamefont {W.}~\bibnamefont {Chen}}, \bibinfo {author}
  {\bibfnamefont {P.}~\bibnamefont {Zhang}}, \ and\ \bibinfo {author}
  {\bibfnamefont {H.}~\bibnamefont {Zhai}},\ }\href {\doibase
  10.1103/PhysRevA.93.043601} {\bibfield  {journal} {\bibinfo  {journal} {Phys.
  Rev. A}\ }\textbf {\bibinfo {volume} {93}},\ \bibinfo {pages} {043601}
  (\bibinfo {year} {2016})}\BibitemShut {NoStop}%
\bibitem [{\citenamefont {Kan\'asz-Nagy}\ \emph {et~al.}(2018)\citenamefont
  {Kan\'asz-Nagy}, \citenamefont {Ashida}, \citenamefont {Shi}, \citenamefont
  {Moca}, \citenamefont {Ikeda}, \citenamefont {F\"olling}, \citenamefont
  {Cirac}, \citenamefont {Zar\'and},\ and\ \citenamefont
  {Demler}}]{Kanasz2018}%
  \BibitemOpen
  \bibfield  {author} {\bibinfo {author} {\bibfnamefont {M.}~\bibnamefont
  {Kan\'asz-Nagy}}, \bibinfo {author} {\bibfnamefont {Y.}~\bibnamefont
  {Ashida}}, \bibinfo {author} {\bibfnamefont {T.}~\bibnamefont {Shi}},
  \bibinfo {author} {\bibfnamefont {C.P.}~\ \bibnamefont
  {Moca}}, \bibinfo {author} {\bibfnamefont {T.~N.}\ \bibnamefont {Ikeda}},
  \bibinfo {author} {\bibfnamefont {S.}~\bibnamefont {F\"olling}}, \bibinfo
  {author} {\bibfnamefont {J.~I.}\ \bibnamefont {Cirac}}, \bibinfo {author}
  {\bibfnamefont {G.}~\bibnamefont {Zar\'and}}, \ and\ \bibinfo {author}
  {\bibfnamefont {E.~A.}\ \bibnamefont {Demler}},\ }\href {\doibase
  10.1103/PhysRevB.97.155156} {\bibfield  {journal} {\bibinfo  {journal} {Phys.
  Rev. B}\ }\textbf {\bibinfo {volume} {97}},\ \bibinfo {pages} {155156}
  (\bibinfo {year} {2018})}\BibitemShut {NoStop}%
\bibitem [{\citenamefont {Nakagawa}\ \emph {et~al.}(2018)\citenamefont
  {Nakagawa}, \citenamefont {Kawakami},\ and\ \citenamefont
  {Ueda}}]{Nakagawa2018}%
  \BibitemOpen
  \bibfield  {author} {\bibinfo {author} {\bibfnamefont {M.}~\bibnamefont
  {Nakagawa}}, \bibinfo {author} {\bibfnamefont {N.}~\bibnamefont {Kawakami}},
  \ and\ \bibinfo {author} {\bibfnamefont {M.}~\bibnamefont {Ueda}},\ }\href
  {\doibase 10.1103/PhysRevLett.121.203001} {\bibfield  {journal} {\bibinfo
  {journal} {Phys. Rev. Lett.}\ }\textbf {\bibinfo {volume} {121}},\ \bibinfo
  {pages} {203001} (\bibinfo {year} {2018})}\BibitemShut {NoStop}%
\bibitem [{\citenamefont {Kuzmenko}\ \emph {et~al.}(2018)\citenamefont
  {Kuzmenko}, \citenamefont {Kuzmenko}, \citenamefont {Avishai},\ and\
  \citenamefont {Jo}}]{Kuzmenko2018}%
  \BibitemOpen
  \bibfield  {author} {\bibinfo {author} {\bibfnamefont {I.}~\bibnamefont
  {Kuzmenko}}, \bibinfo {author} {\bibfnamefont {T.}~\bibnamefont {Kuzmenko}},
  \bibinfo {author} {\bibfnamefont {Y.}~\bibnamefont {Avishai}}, \ and\
  \bibinfo {author} {\bibfnamefont {G.-B.}\ \bibnamefont {Jo}},\ }\href
  {\doibase 10.1103/PhysRevB.97.075124} {\bibfield  {journal} {\bibinfo
  {journal} {Phys. Rev. B}\ }\textbf {\bibinfo {volume} {97}},\ \bibinfo
  {pages} {075124} (\bibinfo {year} {2018})}\BibitemShut {NoStop}%
\bibitem [{\citenamefont {Goto}\ and\ \citenamefont
  {Danshita}(2019)}]{Goto2019}%
  \BibitemOpen
  \bibfield  {author} {\bibinfo {author} {\bibfnamefont {S.}~\bibnamefont
  {Goto}}\ and\ \bibinfo {author} {\bibfnamefont {I.}~\bibnamefont
  {Danshita}},\ }\href {\doibase 10.1103/PhysRevLett.123.143002} {\bibfield
  {journal} {\bibinfo  {journal} {Phys. Rev. Lett.}\ }\textbf {\bibinfo
  {volume} {123}},\ \bibinfo {pages} {143002} (\bibinfo {year}
  {2019})}\BibitemShut {NoStop}%
\bibitem [{\citenamefont {Cappellini}\ \emph {et~al.}(2014)\citenamefont
  {Cappellini}, \citenamefont {Mancini}, \citenamefont {Pagano}, \citenamefont
  {Lombardi}, \citenamefont {Livi}, \citenamefont {Siciliani~de Cumis},
  \citenamefont {Cancio}, \citenamefont {Pizzocaro}, \citenamefont {Calonico},
  \citenamefont {Levi}, \citenamefont {Sias}, \citenamefont {Catani},
  \citenamefont {Inguscio},\ and\ \citenamefont {Fallani}}]{Cappellini2014}%
  \BibitemOpen
  \bibfield  {author} {\bibinfo {author} {\bibfnamefont {G.}~\bibnamefont
  {Cappellini}}, \bibinfo {author} {\bibfnamefont {M.}~\bibnamefont {Mancini}},
  \bibinfo {author} {\bibfnamefont {G.}~\bibnamefont {Pagano}}, \bibinfo
  {author} {\bibfnamefont {P.}~\bibnamefont {Lombardi}}, \bibinfo {author}
  {\bibfnamefont {L.}~\bibnamefont {Livi}}, \bibinfo {author} {\bibfnamefont
  {M.}~\bibnamefont {Siciliani~de Cumis}}, \bibinfo {author} {\bibfnamefont
  {P.}~\bibnamefont {Cancio}}, \bibinfo {author} {\bibfnamefont
  {M.}~\bibnamefont {Pizzocaro}}, \bibinfo {author} {\bibfnamefont
  {D.}~\bibnamefont {Calonico}}, \bibinfo {author} {\bibfnamefont
  {F.}~\bibnamefont {Levi}}, \bibinfo {author} {\bibfnamefont {C.}~\bibnamefont
  {Sias}}, \bibinfo {author} {\bibfnamefont {J.}~\bibnamefont {Catani}},
  \bibinfo {author} {\bibfnamefont {M.}~\bibnamefont {Inguscio}}, \ and\
  \bibinfo {author} {\bibfnamefont {L.}~\bibnamefont {Fallani}},\ }\href
  {\doibase 10.1103/PhysRevLett.113.120402} {\bibfield  {journal} {\bibinfo
  {journal} {Phys. Rev. Lett.}\ }\textbf {\bibinfo {volume} {113}},\ \bibinfo
  {pages} {120402} (\bibinfo {year} {2014})}\BibitemShut {NoStop}%
\bibitem [{\citenamefont {Scazza}\ \emph {et~al.}(2014)\citenamefont {Scazza},
  \citenamefont {Hofrichter}, \citenamefont {H{\"o}fer}, \citenamefont
  {De~Groot}, \citenamefont {Bloch},\ and\ \citenamefont
  {F{\"o}lling}}]{Scazza2014}%
  \BibitemOpen
  \bibfield  {author} {\bibinfo {author} {\bibfnamefont {F.}~\bibnamefont
  {Scazza}}, \bibinfo {author} {\bibfnamefont {C.}~\bibnamefont {Hofrichter}},
  \bibinfo {author} {\bibfnamefont {M.}~\bibnamefont {H{\"o}fer}}, \bibinfo
  {author} {\bibfnamefont {P.~C.}\ \bibnamefont {De~Groot}}, \bibinfo {author}
  {\bibfnamefont {I.}~\bibnamefont {Bloch}}, \ and\ \bibinfo {author}
  {\bibfnamefont {S.}~\bibnamefont {F{\"o}lling}},\ }\href {\doibase
  10.1038/nphys3061} {\bibfield  {journal} {\bibinfo  {journal} {Nature
  Physics}\ }\textbf {\bibinfo {volume} {10}},\ \bibinfo {pages} {779}
  (\bibinfo {year} {2014})}\BibitemShut {NoStop}%
\bibitem [{\citenamefont {Zhang}\ \emph {et~al.}(2014)\citenamefont {Zhang},
  \citenamefont {Bishof}, \citenamefont {Bromley}, \citenamefont {Kraus},
  \citenamefont {Safronova}, \citenamefont {Zoller}, \citenamefont {Rey},\ and\
  \citenamefont {Ye}}]{Zhang2014}%
  \BibitemOpen
  \bibfield  {author} {\bibinfo {author} {\bibfnamefont {X.}~\bibnamefont
  {Zhang}}, \bibinfo {author} {\bibfnamefont {M.}~\bibnamefont {Bishof}},
  \bibinfo {author} {\bibfnamefont {S.~L.}\ \bibnamefont {Bromley}}, \bibinfo
  {author} {\bibfnamefont {C.~V.}\ \bibnamefont {Kraus}}, \bibinfo {author}
  {\bibfnamefont {M.~S.}\ \bibnamefont {Safronova}}, \bibinfo {author}
  {\bibfnamefont {P.}~\bibnamefont {Zoller}}, \bibinfo {author} {\bibfnamefont
  {A.~M.}\ \bibnamefont {Rey}}, \ and\ \bibinfo {author} {\bibfnamefont
  {J.}~\bibnamefont {Ye}},\ }\href {\doibase 10.1126/science.1254978}
  {\bibfield  {journal} {\bibinfo  {journal} {Science}\ }\textbf {\bibinfo
  {volume} {345}},\ \bibinfo {pages} {1467} (\bibinfo {year}
  {2014})}\BibitemShut {NoStop}%
\bibitem [{\citenamefont {Ono}\ \emph {et~al.}(2019)\citenamefont {Ono},
  \citenamefont {Kobayashi}, \citenamefont {Amano}, \citenamefont {Sato},\ and\
  \citenamefont {Takahashi}}]{Ono2019}%
  \BibitemOpen
  \bibfield  {author} {\bibinfo {author} {\bibfnamefont {K.}~\bibnamefont
  {Ono}}, \bibinfo {author} {\bibfnamefont {J.}~\bibnamefont {Kobayashi}},
  \bibinfo {author} {\bibfnamefont {Y.}~\bibnamefont {Amano}}, \bibinfo
  {author} {\bibfnamefont {K.}~\bibnamefont {Sato}}, \ and\ \bibinfo {author}
  {\bibfnamefont {Y.}~\bibnamefont {Takahashi}},\ }\href {\doibase
  10.1103/PhysRevA.99.032707} {\bibfield  {journal} {\bibinfo  {journal} {Phys.
  Rev. A}\ }\textbf {\bibinfo {volume} {99}},\ \bibinfo {pages} {032707}
  (\bibinfo {year} {2019})}\BibitemShut {NoStop}%
\bibitem [{\citenamefont {Riegger}\ \emph {et~al.}(2018)\citenamefont
  {Riegger}, \citenamefont {Darkwah~Oppong}, \citenamefont {H\"ofer},
  \citenamefont {Fernandes}, \citenamefont {Bloch},\ and\ \citenamefont
  {F\"olling}}]{Riegger2018}%
  \BibitemOpen
  \bibfield  {author} {\bibinfo {author} {\bibfnamefont {L.}~\bibnamefont
  {Riegger}}, \bibinfo {author} {\bibfnamefont {N.}~\bibnamefont
  {Darkwah~Oppong}}, \bibinfo {author} {\bibfnamefont {M.}~\bibnamefont
  {H\"ofer}}, \bibinfo {author} {\bibfnamefont {D.~R.}\ \bibnamefont
  {Fernandes}}, \bibinfo {author} {\bibfnamefont {I.}~\bibnamefont {Bloch}}, \
  and\ \bibinfo {author} {\bibfnamefont {S.}~\bibnamefont {F\"olling}},\ }\href
  {\doibase 10.1103/PhysRevLett.120.143601} {\bibfield  {journal} {\bibinfo
  {journal} {Phys. Rev. Lett.}\ }\textbf {\bibinfo {volume} {120}},\ \bibinfo
  {pages} {143601} (\bibinfo {year} {2018})}\BibitemShut {NoStop}%
\bibitem [{\citenamefont {Denschlag}\ \emph {et~al.}(2002)\citenamefont
  {Denschlag}, \citenamefont {Simsarian}, \citenamefont {Häffner},
  \citenamefont {McKenzie}, \citenamefont {Browaeys}, \citenamefont {Cho},
  \citenamefont {Helmerson}, \citenamefont {Rolston},\ and\ \citenamefont
  {Phillips}}]{Denschlag2002}%
  \BibitemOpen
  \bibfield  {author} {\bibinfo {author} {\bibfnamefont {J.~H.}\ \bibnamefont
  {Denschlag}}, \bibinfo {author} {\bibfnamefont {J.~E.}\ \bibnamefont
  {Simsarian}}, \bibinfo {author} {\bibfnamefont {H.}~\bibnamefont {Häffner}},
  \bibinfo {author} {\bibfnamefont {C.}~\bibnamefont {McKenzie}}, \bibinfo
  {author} {\bibfnamefont {A.}~\bibnamefont {Browaeys}}, \bibinfo {author}
  {\bibfnamefont {D.}~\bibnamefont {Cho}}, \bibinfo {author} {\bibfnamefont
  {K.}~\bibnamefont {Helmerson}}, \bibinfo {author} {\bibfnamefont {S.~L.}\
  \bibnamefont {Rolston}}, \ and\ \bibinfo {author} {\bibfnamefont {W.~D.}\
  \bibnamefont {Phillips}},\ }\href {\doibase 10.1088/0953-4075/35/14/307}
  {\bibfield  {journal} {\bibinfo  {journal} {Journal of Physics B}\ }\textbf {\bibinfo {volume} {35}},\ \bibinfo
  {pages} {3095} (\bibinfo {year} {2002})}\BibitemShut {NoStop}%
\bibitem [{\citenamefont {Takata}\ \emph {et~al.}(2019)\citenamefont {Takata},
  \citenamefont {Nakajima}, \citenamefont {Kobayashi}, \citenamefont {Ono},
  \citenamefont {Amano},\ and\ \citenamefont {Takahashi}}]{Takata2019}%
  \BibitemOpen
  \bibfield  {author} {\bibinfo {author} {\bibfnamefont {Y.}~\bibnamefont
  {Takata}}, \bibinfo {author} {\bibfnamefont {S.}~\bibnamefont {Nakajima}},
  \bibinfo {author} {\bibfnamefont {J.}~\bibnamefont {Kobayashi}}, \bibinfo
  {author} {\bibfnamefont {K.}~\bibnamefont {Ono}}, \bibinfo {author}
  {\bibfnamefont {Y.}~\bibnamefont {Amano}}, \ and\ \bibinfo {author}
  {\bibfnamefont {Y.}~\bibnamefont {Takahashi}},\ }\href {\doibase
  10.1063/1.5110037} {\bibfield  {journal} {\bibinfo  {journal} {Review of
  Scientific Instruments}\ }\textbf {\bibinfo {volume} {90}},\ \bibinfo {pages}
  {083002} (\bibinfo {year} {2019})}\BibitemShut {NoStop}%
\bibitem [{\citenamefont {Taie}\ \emph {et~al.}(2010)\citenamefont {Taie},
  \citenamefont {Takasu}, \citenamefont {Sugawa}, \citenamefont {Yamazaki},
  \citenamefont {Tsujimoto}, \citenamefont {Murakami},\ and\ \citenamefont
  {Takahashi}}]{Taie2010}%
  \BibitemOpen
  \bibfield  {author} {\bibinfo {author} {\bibfnamefont {S.}~\bibnamefont
  {Taie}}, \bibinfo {author} {\bibfnamefont {Y.}~\bibnamefont {Takasu}},
  \bibinfo {author} {\bibfnamefont {S.}~\bibnamefont {Sugawa}}, \bibinfo
  {author} {\bibfnamefont {R.}~\bibnamefont {Yamazaki}}, \bibinfo {author}
  {\bibfnamefont {T.}~\bibnamefont {Tsujimoto}}, \bibinfo {author}
  {\bibfnamefont {R.}~\bibnamefont {Murakami}}, \ and\ \bibinfo {author}
  {\bibfnamefont {Y.}~\bibnamefont {Takahashi}},\ }\href {\doibase
  10.1103/PhysRevLett.105.190401} {\bibfield  {journal} {\bibinfo  {journal}
  {Phys. Rev. Lett.}\ }\textbf {\bibinfo {volume} {105}},\ \bibinfo {pages}
  {190401} (\bibinfo {year} {2010})}\BibitemShut {NoStop}%
\bibitem [{\citenamefont {Bettermann}\ \emph {et~al.}()\citenamefont
  {Bettermann}, \citenamefont {Oppong}, \citenamefont {Pasqualetti},
  \citenamefont {Riegger}, \citenamefont {Bloch},\ and\ \citenamefont
  {Fölling}}]{Bettermann2020}%
  \BibitemOpen
  \bibfield  {author} {\bibinfo {author} {\bibfnamefont {O.}~\bibnamefont
  {Bettermann}}, \bibinfo {author} {\bibfnamefont {N.~D.}\ \bibnamefont
  {Oppong}}, \bibinfo {author} {\bibfnamefont {G.}~\bibnamefont {Pasqualetti}},
  \bibinfo {author} {\bibfnamefont {L.}~\bibnamefont {Riegger}}, \bibinfo
  {author} {\bibfnamefont {I.}~\bibnamefont {Bloch}}, \ and\ \bibinfo {author}
  {\bibfnamefont {S.}~\bibnamefont {Fölling}},\ }\href@noop {} {}\Eprint
  {http://arxiv.org/abs/2003.10599} {arXiv:2003.10599 [cond-mat.quant-gas]}
  \BibitemShut {NoStop}%
\bibitem [{\citenamefont {Syassen}\ \emph {et~al.}(2008)\citenamefont
  {Syassen}, \citenamefont {Bauer}, \citenamefont {Lettner}, \citenamefont
  {Volz}, \citenamefont {Dietze}, \citenamefont {Garc{\'\i}a-Ripoll},
  \citenamefont {Cirac}, \citenamefont {Rempe},\ and\ \citenamefont
  {D{\"u}rr}}]{Syassen2008}%
  \BibitemOpen
  \bibfield  {author} {\bibinfo {author} {\bibfnamefont {N.}~\bibnamefont
  {Syassen}}, \bibinfo {author} {\bibfnamefont {D.~M.}\ \bibnamefont {Bauer}},
  \bibinfo {author} {\bibfnamefont {M.}~\bibnamefont {Lettner}}, \bibinfo
  {author} {\bibfnamefont {T.}~\bibnamefont {Volz}}, \bibinfo {author}
  {\bibfnamefont {D.}~\bibnamefont {Dietze}}, \bibinfo {author} {\bibfnamefont
  {J.~J.}\ \bibnamefont {Garc{\'\i}a-Ripoll}}, \bibinfo {author} {\bibfnamefont
  {J.~I.}\ \bibnamefont {Cirac}}, \bibinfo {author} {\bibfnamefont
  {G.}~\bibnamefont {Rempe}}, \ and\ \bibinfo {author} {\bibfnamefont
  {S.}~\bibnamefont {D{\"u}rr}},\ }\href {\doibase 10.1126/science.1155309}
  {\bibfield  {journal} {\bibinfo  {journal} {Science}\ }\textbf {\bibinfo
  {volume} {320}},\ \bibinfo {pages} {1329} (\bibinfo {year}
  {2008})}\BibitemShut {NoStop}%
\bibitem [{\citenamefont {Tomita}\ \emph {et~al.}(2019)\citenamefont {Tomita},
  \citenamefont {Nakajima}, \citenamefont {Takasu},\ and\ \citenamefont
  {Takahashi}}]{Tomita2019}%
  \BibitemOpen
  \bibfield  {author} {\bibinfo {author} {\bibfnamefont {T.}~\bibnamefont
  {Tomita}}, \bibinfo {author} {\bibfnamefont {S.}~\bibnamefont {Nakajima}},
  \bibinfo {author} {\bibfnamefont {Y.}~\bibnamefont {Takasu}}, \ and\ \bibinfo
  {author} {\bibfnamefont {Y.}~\bibnamefont {Takahashi}},\ }\href {\doibase
  10.1103/PhysRevA.99.031601} {\bibfield  {journal} {\bibinfo  {journal} {Phys.
  Rev. A}\ }\textbf {\bibinfo {volume} {99}},\ \bibinfo {pages} {031601(R)}
  (\bibinfo {year} {2019})}\BibitemShut {NoStop}%
\bibitem [{\citenamefont {Tsunetsugu}\ \emph {et~al.}(1997)\citenamefont
  {Tsunetsugu}, \citenamefont {Sigrist},\ and\ \citenamefont
  {Ueda}}]{Tsunetsugu1997}%
  \BibitemOpen
  \bibfield  {author} {\bibinfo {author} {\bibfnamefont {H.}~\bibnamefont
  {Tsunetsugu}}, \bibinfo {author} {\bibfnamefont {M.}~\bibnamefont {Sigrist}},
  \ and\ \bibinfo {author} {\bibfnamefont {K.}~\bibnamefont {Ueda}},\ }\href
  {\doibase 10.1103/RevModPhys.69.809} {\bibfield  {journal} {\bibinfo
  {journal} {Rev. Mod. Phys.}\ }\textbf {\bibinfo {volume} {69}},\ \bibinfo
  {pages} {809} (\bibinfo {year} {1997})}\BibitemShut {NoStop}%
\bibitem [{\citenamefont {Cheng}\ \emph {et~al.}(2017)\citenamefont {Cheng},
  \citenamefont {Zhang}, \citenamefont {Zhang},\ and\ \citenamefont
  {Zhai}}]{Cheng2017}%
  \BibitemOpen
  \bibfield  {author} {\bibinfo {author} {\bibfnamefont {Y.}~\bibnamefont
  {Cheng}}, \bibinfo {author} {\bibfnamefont {R.}~\bibnamefont {Zhang}},
  \bibinfo {author} {\bibfnamefont {P.}~\bibnamefont {Zhang}}, \ and\ \bibinfo
  {author} {\bibfnamefont {H.}~\bibnamefont {Zhai}},\ }\href {\doibase
  10.1103/PhysRevA.96.063605} {\bibfield  {journal} {\bibinfo  {journal} {Phys.
  Rev. A}\ }\textbf {\bibinfo {volume} {96}},\ \bibinfo {pages} {063605}
  (\bibinfo {year} {2017})}\BibitemShut {NoStop}%
\bibitem [{\citenamefont {Zhang}\ and\ \citenamefont
  {Zhang}(2018)}]{Zhang2018}%
  \BibitemOpen
  \bibfield  {author} {\bibinfo {author} {\bibfnamefont {R.}~\bibnamefont
  {Zhang}}\ and\ \bibinfo {author} {\bibfnamefont {P.}~\bibnamefont {Zhang}},\
  }\href {\doibase 10.1103/PhysRevA.98.043627} {\bibfield  {journal} {\bibinfo
  {journal} {Phys. Rev. A}\ }\textbf {\bibinfo {volume} {98}},\ \bibinfo
  {pages} {043627} (\bibinfo {year} {2018})}\BibitemShut {NoStop}%
\bibitem [{\citenamefont {Goto}\ and\ \citenamefont
  {Danshita}()}]{GotoPrivCom}%
  \BibitemOpen
  \bibfield  {author} {\bibinfo {author} {\bibfnamefont {S.}~\bibnamefont
  {Goto}}\ and\ \bibinfo {author} {\bibfnamefont {I.}~\bibnamefont
  {Danshita}},\ }\href@noop {} {\bibinfo  {journal} {private communication}\
  }\BibitemShut {NoStop}%
\bibitem [{\citenamefont {Yamamoto}\ \emph {et~al.}(2016)\citenamefont
  {Yamamoto}, \citenamefont {Kobayashi}, \citenamefont {Kuno}, \citenamefont
  {Kato},\ and\ \citenamefont {Takahashi}}]{Yamamoto2016}%
  \BibitemOpen
\bibfield  {journal} {  }\bibfield  {author} {\bibinfo {author} {\bibfnamefont
  {R.}~\bibnamefont {Yamamoto}}, \bibinfo {author} {\bibfnamefont
  {J.}~\bibnamefont {Kobayashi}}, \bibinfo {author} {\bibfnamefont
  {T.}~\bibnamefont {Kuno}}, \bibinfo {author} {\bibfnamefont {K.}~\bibnamefont
  {Kato}}, \ and\ \bibinfo {author} {\bibfnamefont {Y.}~\bibnamefont
  {Takahashi}},\ }\href {\doibase 10.1088/1367-2630/18/2/023016} {\bibfield
  {journal} {\bibinfo  {journal} {New Journal of Physics}\ }\textbf {\bibinfo
  {volume} {18}},\ \bibinfo {pages} {023016} (\bibinfo {year}
  {2016})}\BibitemShut {NoStop}%
\bibitem [{\citenamefont {Miranda}\ \emph {et~al.}(2015)\citenamefont
  {Miranda}, \citenamefont {Inoue}, \citenamefont {Okuyama}, \citenamefont
  {Nakamoto},\ and\ \citenamefont {Kozuma}}]{Miranda2015}%
  \BibitemOpen
  \bibfield  {author} {\bibinfo {author} {\bibfnamefont {M.}~\bibnamefont
  {Miranda}}, \bibinfo {author} {\bibfnamefont {R.}~\bibnamefont {Inoue}},
  \bibinfo {author} {\bibfnamefont {Y.}~\bibnamefont {Okuyama}}, \bibinfo
  {author} {\bibfnamefont {A.}~\bibnamefont {Nakamoto}}, \ and\ \bibinfo
  {author} {\bibfnamefont {M.}~\bibnamefont {Kozuma}},\ }\href {\doibase
  10.1103/PhysRevA.91.063414} {\bibfield  {journal} {\bibinfo  {journal} {Phys.
  Rev. A}\ }\textbf {\bibinfo {volume} {91}},\ \bibinfo {pages} {063414}
  (\bibinfo {year} {2015})}\BibitemShut {NoStop}%
\bibitem [{\citenamefont {V.~Borzenets}\ \emph {et~al.}(2020)\citenamefont
  {V.~Borzenets}, \citenamefont {Shim}, \citenamefont {Chen}, \citenamefont
  {Ludwig}, \citenamefont {Wieck}, \citenamefont {Tarucha}, \citenamefont
  {Sim},\ and\ \citenamefont {Yamamoto}}]{Borzenets2020}%
  \BibitemOpen
  \bibfield  {author} {\bibinfo {author} {\bibfnamefont {I.}~\bibnamefont
  {V.~Borzenets}}, \bibinfo {author} {\bibfnamefont {J.}~\bibnamefont {Shim}},
  \bibinfo {author} {\bibfnamefont {J.~C.~H.}\ \bibnamefont {Chen}}, \bibinfo
  {author} {\bibfnamefont {A.}~\bibnamefont {Ludwig}}, \bibinfo {author}
  {\bibfnamefont {A.~D.}\ \bibnamefont {Wieck}}, \bibinfo {author}
  {\bibfnamefont {S.}~\bibnamefont {Tarucha}}, \bibinfo {author} {\bibfnamefont
  {H.-S.}\ \bibnamefont {Sim}}, \ and\ \bibinfo {author} {\bibfnamefont
  {M.}~\bibnamefont {Yamamoto}},\ }\href {\doibase 10.1038/s41586-020-2058-6}
  {\bibfield  {journal} {\bibinfo  {journal} {Nature}\ }\textbf {\bibinfo
  {volume} {579}},\ \bibinfo {pages} {210} (\bibinfo {year}
  {2020})}\BibitemShut {NoStop}%
\end{thebibliography}
\end{document}